\documentclass[pre,twocolumn,amsmath,amssymb,nofootinbib,floatfix,superscriptaddress]{revtex4}

\usepackage{graphicx,bm,float,physics}
\usepackage{dsfont}
\usepackage{xcolor}
\usepackage[normalem]{ulem}

\begin{document}

\title{O(5) multicriticality in the 3D two flavor SU(2) lattice gauge Higgs model}

\author{Claudio Bonati} 
\affiliation{Dipartimento di Fisica dell'Universit\`a di Pisa and 
INFN Sezione di Pisa, Largo Pontecorvo 3, I-56127 Pisa, Italy}

\author{Ivan Soler Calero} 
\affiliation{Dipartimento di Fisica dell'Universit\`a di Pisa and 
INFN Sezione di Pisa, Largo Pontecorvo 3, I-56127 Pisa, Italy}

\date{\today}

\begin{abstract}
We numerically investigate the multicritical behavior of the three dimensional
lattice system in which a SU(2) gauge field is coupled to two flavors of scalar
fields transforming in the fundamental representation of the gauge group. In
this system a multicritical point is present, where the global symmetry
O(2)$\oplus$O(3) gets enlarged to O(5). Such a symmetry enlargement is hindered
for generic systems by the instability of the O(5) multicritical point, but the
SU(2) gauge symmetry prevents the appearance of the term triggering the
instability. All the numerical results obtained in this lattice gauge model
fully support the expectations coming from the O(2)$\oplus$O(3) multicritical
Landau-Ginzburg-Wilson $\phi^4$ theory, and we discuss possible implications of
these results for some models of deconfined quantum criticality.
\end{abstract}

\maketitle

\section{Introduction}
\label{intro}

The behavior of physical systems close to a continuous phase transition is very
peculiar: due to the divergence of the correlation length, universal scaling
laws emerge, which are largely independent of the microscopic details of the
systems~\cite{Landau-book, Stanley-book}.  The theoretical framework needed to
understand this behavior is the renormalization group (RG)~\cite{WK-74,
Fisher-75, PV-02, ZJ-book}, and for most of the systems of physical interest
the critical point can be reached by tuning just two control parameters.  The
tuning of a larger number of control parameters is however needed to approach
peculiar critical points, known as multicritical points, which
can be found in quantum lattice gas and antiferromagnetic
systems~\cite{LF-72, FN-74, NKF-74, KNF-76}, in anisotropic magnetic systems 
\cite{BA-75, DF-77, CHKetal-82, ZCD-21}, and in fluid mixtures
\cite{BEG-71, Griffiths-73, WR-73, SAS-75, LS-DL, KS-DL}. 

A typical reason for the appearance of multicritical phenomena in a physical
system is the presence of different order parameters, related to independent
symmetries, with the multicritical point being located where different critical
lines meet. Depending on the number of critical lines which cross at the
multicritical point, one speaks of bicritical, tricritical, tetracritical
points, and so on. Particularly interesting are those cases in which the
effective symmetry (i.~e.  the symmetry of large distance critical
correlations) at the multicritical point is larger than the one displayed
by the original model for generic values of the control parameters.

The possibility of a critical symmetry enlargement at multicritical points has
been already recognized in the seminal paper \cite{NKF-74}, where it was shown,
using the $\varepsilon$-expansion up to $O(\varepsilon^2)$ and a Pad\'e
resummation, that a three dimensional model with two independent Ising order
parameters (i.~e. a symmetry $\mathbb{Z}_2\oplus\mathbb{Z}_2$) can exhibit a
bicritical point with O(2) symmetry. It is however important to note that the
operatorial content of the multicritical O(2) theory is not the same of
the critical O(2) theory, since the symmetry constraining the RG flow is
different in the two cases, and in particular the scaling dimension of the
first irrelevant operator is different in the two cases, see, e.~g., \cite{CPV-03}.

A three dimensional O(5) multicritical behavior, emerging when a 2-component
superconducting order parameter interacts with a 3-component antiferromagnetic
order parameter (corresponding to the symmetry
$\mathrm{O}(2)\oplus\mathrm{O}(3)$), has been suggested as a possible mechanism
for high-$T$ superconductivity \cite{Zhang-97, AH-99, ZHAHA-99, AH-00, DHZ-04},
however it was later shown that such an O(5) multicritical point is in fact
unstable, the only stable multicritical point being in this case the decoupled
one \cite{Aharony-02, CPV-02-comm, CPV-03, PV-05, HPV-05}. This means that the
O(5) symmetry enlargement can not generically happen without the tuning of a
further parameter.  It was however quite recently noted \cite{BFPV-21x} that,
in the SU(2) gauge model with two degenerate flavors of fundamental scalar
fields, the gauge symmetry protects the O(5) symmetry, preventing the RG flow
from generating the term which triggers the instability.

The principal aim of this work is to carefully investigate the multicritical
behavior of the three dimensional lattice SU(2) gauge model with two flavors of
fundamental scalar fields. Indeed in \cite{BFPV-21x} the presence of the O(5)
multicritical behavior has not been thoroughly studied, and only some general
features of the phase diagram of this model have been numerically tested. The
motivation for studying this multicritical point is to investigate the possible
interplay between gauge and global symmetries at a multicritical point. 

Let us explain this point more in detail: three dimensional systems with gauge
symmetries are very common in condensed matter physics \cite{Fradkin_book,
Moessner_book, Wen-book, Sachdev-19}, and their critical behaviors present
several distinctive features, depending mainly on which degrees of
freedom become critical at the transition, see \cite{BPV-24-rev} for a
review. If only matter fields develop long range correlations, two cases are
possible: (i) the critical properties can be described by an effective
Landau-Ginzburg-Wilson (LGW) $\phi^4$ Hamiltonian, displaying the same symmetry
breaking pattern of the original model, and whose order parameter is a
gauge invariant local composite operator (standard LGW case); (ii) the critical
properties can be described by an effective LGW $\phi^4$ Hamiltonian whose
order parameter is not gauge invariant (LGW$^{\times}$ case, sometimes also
denoted by LGW$^*$). 
The case (i) is typical of systems characterized by the presence of a
``small'' number of scalar fields and continuous (Abelian or non-Abelian) gauge
symmetry (e.~g., $N_f\lesssim 10$ in the three dimensional noncompact Abelian case, see
\cite{BPV-24-rev} for more details), while the case (ii) is usually observed in
lattice systems displaying continuous global symmetries and discrete gauge
symmetry, the most studied case being that of the $\mathbb{Z}_2$ gauge group.
If instead gauge degrees of freedom are critical we can
have the following cases: (iii) if matter fields are also critical, the
effective Hamiltonian has to be a gauge field theory (GFT case); (iv) if matter
fields do not develop long range order, we have a topological transition,
driven by topological gauge modes, and characterized by the absence of local
gauge-invariant order parameters (topological transition).
The case (iii) is often found in models characterized by continuous
gauge symmetries and a ``large'' number of scalar fields (e.~g., $N_f\gtrsim
10$ in the three dimensional noncompact Abelian case, see \cite{BPV-24-rev} for
more details) , while the topological transition case (iv) is usually
encountered when discrete Abelian gauge groups are present. Note that this
classification refers to transitions (or transitions lines) and not to physical
models: it is quite common for different transitions, characterized by
different critical behaviors, to be present in the phase diagram of a given
model.

The critical transitions of the three dimensional SU(2) model with a small
number of flavors $N_f$ have been shown to be described by the standard LGW paradigm
\cite{BPV-19a, BPV-19b, BFPV-21x}, while for $N_f\gtrsim 30$ transitions of the
GFT case are also present \cite{BFPV-21x, BPSV-24}. The question we want to
answer in this work is thus the following: in a gauge theory in which several
critical lines exist, associated to critical behaviors described by the LGW
gauge invariant paradigm, are the eventual multicritical points also described
by the corresponding multicritical gauge-invariant LGW effective theory? This
point does not seem to have been carefully investigated in the literature, also
because the numerical study of multicritical points is notoriously complicated,
typically requiring the fine tuning of several independent parameters. 

To the best of our knowledge the only case which attracted some attention is
that of the $\mathbb{Z}_2$ gauge-Higgs model. In such a model a line of
$\mathbb{Z}_2$ LGW$^{\times}$ transitions joins a line of $\mathbb{Z}_2$
topological transitions along the self-dual line. The self-duality of the model
greatly simplifies the numerical identification of the multicritical point,
however, since neither of the two transition is a standard LGW transition, is
not clear to what extent standard multicritical theory can be applied. For this
reason the nature of the multicritical point is still debated, despite the fact
that the numerically computed critical exponents nicely match those expected at
a multicritical O(2) point \cite{SSN-21, BPV-22, OKGR-24, BPV-24comm}.

A further reason for studying the O(5) multicriticality of this SU(2) model is
its possible relation with deconfined criticality \cite{SVBSF-04, SBSVF-04,
LS-04}. Indeed several studies of models relevant for deconfined criticality
\cite{TH-05, SF-06, NCSOS-15, NSCOS-15-bis, TS-20, ZHZH-24, DLGL-24, DES-24}
have found evidence of emerging O(5) symmetry, although typically in a quite
indirect way. In particular, an accurate comparison with universal quantities
of the O(5) critical point (see, e.~g., \cite{Hasenbusch-22} for recent
determinations) does not seem to have been performed.  In many cases this is
due to the presence of unusually strong corrections to scaling, which make it
very difficult to even determine the order of the transition \cite{JNCW-08,
Sandvik-10, BDA-10, Kaul-11, ZHDKPS-13, HSOMLWTK-13}.  A careful study of the
SU(2) model with two flavors of scalar fields can thus provide some hints of
what happens in models that are theoretically less understood, suggesting
possible schemes for interpreting the numerically observed phenomenology. We
will come back to this point in the conclusions.

The paper is organized as follows. In Sec.~\ref{lattice} we introduce the two
flavors SU(2) lattice model and describe some of its properties, while in
Sec.~\ref{multi} we summarize some results that have been obtained for the
O($N_1$)$\oplus$O($N_2$) multicritical LGW model, considering in particular the
case $N_1=2, N_2=3$. In Sec.~\ref{results} we discuss our numerical results,
and in Sec.~\ref{concl} we draw our conclusions.

\section{The SU(2) lattice model}
\label{lattice}

The model we consider in this paper is defined by the lattice Hamiltonian
\begin{align}
H =&- 2J\sum_{{\bm x},\mu} {\rm Re}\,{\rm Tr} \,\Phi_{\bm
    x}^\dagger \, U_{{\bm x},\mu} \, \Phi_{{\bm
      x}+\hat{\mu}}^{\phantom\dagger} + {v\over 4} \sum_{\bm x} {\rm
    Tr}\,(\Phi_{\bm x}^\dagger\Phi_{\bm x})^2 \nonumber \\ 
  &-
  \frac{\gamma}{2} \sum_{{\bm x},\mu>\nu} {\rm Re} \, {\rm Tr}\,
  \big(U_{{\bm x},\mu} \,U_{{\bm x}+\hat{\mu},\nu} \,U_{{\bm
        x}+\hat{\nu},\mu}^\dagger \,U_{{\bm x},\nu}^\dagger\big)\ .
\label{ham}
\end{align}
Scalar fields are represented by the complex matrices $\Phi^{af}_{\bm x}$,
where $a=1,2$ is a color index and $f=1,2$ is a flavor index. Scalars are
defined on the lattice sites, they transform in the fundamental representation
of the gauge group, and satisfy the unit-length constraint ${\rm Tr}\,
\Phi_{\bm x}^\dagger \Phi_{\bm x}^{\phantom\dagger} = 1$ (fixed just to avoid
introducing a further parameter in the Hamiltonian). Gauge variables are
instead represented by ${\rm SU}(2)$ color matrices $U_{{\bm x},\mu}$ defined
on the lattice links~\cite{Wilson-74}. 
The last term of the lattice Hamiltonian $H$ is the so called Wilson
action, which is the simplest gauge invariant self-interaction term for gauge
fields on the lattice.  When $\gamma=0$ gauge fields are randomly distributed
in the gauge group if we neglect their coupling to matter (i.~e., for $J=0$),
while in the large $\gamma$ limit the expectation value of the local operator
multiplying $\gamma$ (the so called plaquette) approaches one. When all
plaquettes are equal to one it is possible, in the infinite volume limit, to
perform a gauge fixing which removes the gauge fields from the theory while
maintaining the locality of the Hamiltonian. 
This lattice model is a particular case of that studied in \cite{BFPV-21x,
BPSV-24}, where a generic number of colors and flavors was considered; see also
\cite{BPV-19a, BPV-19b} for the case $v=0$.

We assume the model in Eq.~\eqref{ham} to be defined on a cubic lattice of
linear size $L$, with periodic boundary conditions, and all couplings are
measured in units of the temperature, which is equivalent to fixing
$\beta=1/(k_B T)=1$. We thus write the partition function as 
\begin{equation}
Z = \sum_{\{\Phi,U\}} \exp(- H)\ .
\end{equation}

The Hamiltonian $H$ is invariant under the local symmetry
\begin{equation}
U_{{\bm x},\mu}\to G_{\bm x}U_{{\bm x},\mu}\ ,\quad  
\Phi_{\bm x}\to G_{\bm x}\Phi_{\bm x}\ , 
\end{equation}
with $G_{\bm x}\in \mathrm{SU(2)}$, and under the global continuous
symmetry
\begin{equation}
\Phi_{\bm x}\to \Phi_{\bm x}M\ , 
\end{equation}
with $M\in\mathrm{U(2)}$. The group U(2) is not a simple group, and we will now
introduce two different gauge invariant order parameters, associated to the
global SU(2) symmetry and to the global U(1) symmetry, respectively. Since
$\mathbb{Z}_2$ is the center of the SU(2) group, the Abelian part of the global
symmetry is more properly U(1)/$\mathbb{Z}_2$, and indeed the associated order
parameter is quadratic in the scalar fields.

The order parameter that we use to monitor the spontaneous symmetry breaking of
the SU(2) non-Abelian component of the global symmetry is
$Q_{\bm x}=\Phi_{\bm x}^{\dag}\Phi_{\bm x}-\mathds{1}/2$ or, in components,
\begin{equation}\label{Q}
Q^{fg}_{\bm x}=
\sum_a \bar{\Phi}^{af}_{\bm x}\Phi^{ag}_{\bm x}-\frac{\delta^{fg}}{2}\ ,
\end{equation}
which transforms in the adjoint representation of the SU(2) global symmetry and
is gauge invariant.  The order parameter for the global
$\mathrm{U(1)}/\mathbb{Z}_2$ symmetry is instead
\begin{equation}\label{Y}
Y_{\bm x}^{fg}=\epsilon^{ab}\Phi_{\bm x}^{af}\Phi_{\bm x}^{bg}=
\epsilon^{fg}\det\Phi_{\bm x}\ ,
\end{equation}
which is also obviously gauge invariant.

When $v=0$ the global symmetry of the Hamiltonian in Eq.~\eqref{ham} is in fact
larger: if we define the $2\times 4$ matrix $\Gamma^{a\gamma}_{\bm x}$ by
\begin{equation}
\Gamma_{\bm x}^{a\gamma}=\left\{\begin{array}{ll} 
\Phi_{\bm x}^{a\gamma} & \mathrm{if}\ 1\le \gamma\le 2 \\
\sum_b \epsilon^{ab}\bar{\Phi}_{\bm x}^{b (\gamma-2)} & \mathrm{if}\ 3\le \gamma \le 4\ ,
\end{array}\right.
\end{equation}
which transforms in the fundamental representation of SU(2) gauge group, 
it is not difficult to show (see \cite{BPV-19b} for details) that 
\begin{equation}
{\rm Re}\,{\rm Tr} \,\Phi_{\bm
    x}^\dagger \, U_{{\bm x},\mu} \, \Phi_{{\bm
      x}+\hat{\mu}}^{\phantom\dagger} 
=\frac{1}{2}{\rm Re}\,{\rm Tr} \,\Gamma_{\bm
    x}^\dagger \, U_{{\bm x},\mu} \, \Gamma_{{\bm
      x}+\hat{\mu}}^{\phantom\dagger}\ , 
\end{equation}
and that this term is invariant under the transformation
$\Gamma_{\bm x}\to \Gamma_{\bm x}N$, where $N$ is a SU(4) 
matrix satisfying
\begin{equation}
NJN^T=J\ ,\quad  J=\left(\begin{array}{cc} 0 & -1 \\ 1 & 0\end{array}\right)\ .
\end{equation}
The global invariance group of the $v=0$ model is thus the unitary symplectic
group Sp(2), which is isomorphic (up to a discrete $\mathbb{Z}_2$ subgroup) to
SO(5), see, e.~g., \cite{Simon-book}. 

It is simple to verify by direct computation that $\mathrm{Tr}\Gamma_{\bm
x}^{\dag}\Gamma_{\bm x}=2$, and using $\Gamma_{\bm x}$ we can write an
order parameter which is the analogous of $Q_{\bm x}$ for the SO(5) symmetry:
\begin{equation}
\mathcal{T}_{\bm x}^{\alpha\beta}=\sum_a\bar{\Gamma}_{\bm x}^{a\alpha}\Gamma_{\bm x}^{a\beta}-
\frac{\delta^{\alpha\beta}}{2}\ ,
\end{equation}
with $1\le \alpha,\beta\le 4$. It is not difficult to verify that
\begin{equation}
\mathcal{T}_{\bm x}=\left(\begin{array}{cc} Q_{\bm x} & \bar{Y}_{\bm x} \\ 
-Y_{\bm x} & \bar{Q}_{\bm x} \end{array}\right)\ ,
\end{equation}
moreover, if we define the five real numbers $\phi_{{\bm x},i}$ (with $i=1,\ldots, 5$) by
\begin{equation}
\begin{aligned}
Q_{\bm x}&=\left(\begin{array}{cc} \phi_{{\bm x},3} &  \phi_{{\bm x},1} -i\phi_{{\bm x},2} \\
\phi_{{\bm x},1} + i\phi_{{\bm x},2} & -  \phi_{{\bm x},3} \end{array}\right)\\
Y_{\bm x}&=\left(\begin{array}{cc} 0 &  -\phi_{{\bm x},4} -i\phi_{{\bm x},5} \\
\phi_{{\bm x},4} + i\phi_{{\bm x},5} & 0 \end{array}\right)\ ,
\end{aligned}
\end{equation}
it is possible to verify that under an infinitesimal Sp(2) transformation the
vector ${\bm \phi}_{\bm x}=(\phi_{{\bm x},1},\ldots, \phi_{{\bm x},5})$
transforms in the fundamental representation of SO(5), thus explicitly
realizing the isomorphism $\mathrm{SO(5)}=\mathrm{Sp(2)}/\mathbb{Z}_2$ (see
\cite{Simon-book}). 

When $v\neq 0$ the SO(5) symmetry is explicitly broken,
indeed it is simple to verify that
\begin{equation}\label{trphi2}
\begin{aligned}
\mathrm{Tr}(\Phi_{\bm x}^{\dag}\Phi_{\bm x})^2&=\mathrm{Tr}(Q_{\bm x}^2)-1/2=\\
&=2(\phi_{{\bm x},1}^2+\phi_{{\bm x},2}^2+\phi_{{\bm x},3}^2)-1/2\ ,
\end{aligned}
\end{equation}
and for $v\neq 0$ we thus recover the SO(2)$\oplus$SO(3) symmetry that was
identified before.  For $v>0$ we are introducing a penalty term disfavoring
the spontaneous breaking of the SO(3) symmetry, and we thus expect transitions
in the SO(2) universality class, while for $v<0$ we are effectively disfavoring
the spontaneous breaking of the SO(2) symmetry, and we thus expect transitions in
the SO(3) universality class. These expectations are in fact confirmed by a mean
field analysis, see \cite{BFPV-21x}.

Let us finally note that the Hamiltonian in Eq.~\eqref{ham} is invariant both
under charge conjugation and parity. This implies that the effective Hamiltonian
to be discussed in the following section will in fact be symmetric under
general orthogonal transformations, and not only under the 
special orthogonal ones.

\section{Some results for the O($N_1$)$\oplus$O($N_2$) multicritical theory}
\label{multi}

The most general LGW Hamiltonian density which is invariant under the symmetry
O($N_1$)$\oplus$O($N_2$) and contains up to quartic terms can be written in the
form \cite{NKF-74, CPV-03, HPV-05}
\begin{equation}
\begin{aligned}
\mathcal{H}=\frac{1}{2}[(\partial_{\mu} \psi_1)^2+(\partial_{\mu}\psi_2)^2]+\frac{1}{2}[r_1\psi_1^2+r_2\psi_2^2]\\
+\frac{1}{4!}[u_1(\psi_1^2)^2+u_2(\psi_2^2)^2+2w\psi_1^2\psi_2^2]\ ,
\end{aligned}
\end{equation}
where the fields $\psi_1$ and $\psi_2$ have $N_1$ and $N_2$ components,
respectively. When $r_1$ and $r_2$ are set to their critical values, this
Hamiltonian describes the multicritical behavior (if it exists at all) of a
system in which a O($N_1$) order parameter interacts with an independent
O($N_2$) order parameter.

A possible strategy to understand whether the multicritical behavior can
display an enlarged O($N$) symmetry, with $N=N_1+N_2$, is to study the scaling
dimensions of the perturbations with O($N_1$)$\oplus$O($N_2$) symmetry of the
O($N$) critical point. These perturbations can be expanded \cite{Wegner-72} as
linear combinations of homogeneous polynomials of degree $m$ which transforms
according to the spin $\ell$ (with $m\ge \ell$) representation of O($N$), see,
e.~g., \cite{CPV-03} for explicit examples.  Since polynomial with different
spin values do not mix under RG transformation, this basis is particularly convenient
to classify the relevant perturbations of the O($N$) critical theory. 

Close to four dimensions one expects the possible relevant perturbations to be
those with $m,\ell\le 4$, and using the O($N_1$)$\oplus$O($N_2$) symmetry it is
possible to show that only 5 such terms $\mathcal{P}_{m\ell}$ exist:
$\mathcal{P}_{2,0}$, $\mathcal{P}_{2,2}$, $\mathcal{P}_{4,0}$,
$\mathcal{P}_{4,2}$, and $\mathcal{P}_{4,4}$ (see \cite{CPV-03} for more
details).  The $\ell=0$ perturbations correspond to O($N$) invariant
operators, and in particular the RG scaling dimension $y_{2,0}$ of
$\mathcal{P}_{2,0}$ is related to the correlation length critical exponent of
the O$(N)$ model: 
\begin{equation}
y_{2,0}=1/\nu_{\mathrm{O(}N\mathrm{)}}\ .
\end{equation}
Analogously the RG scaling dimension $y_{4,0}$ of $\mathcal{P}_{4,0}$ is related
to the leading scaling correction exponent of the O($N$) theory by 
\begin{equation}
y_{4,0}=-\omega_{\mathrm{O(}N\mathrm{)}}\ .
\end{equation}
The RG dimension $y_{2,2}$ of the spin two quadratic perturbation
$\mathcal{P}_{2,2}$ gives (when the multicritical behavior exists) the second
correlation length critical exponent
\begin{equation}
y_{2,2}=1/\nu'\ ,
\end{equation}
and determines also the crossover exponent
\begin{equation}
\phi_T=y_{2,2}\nu_{\mathrm{O(}N\mathrm{)}}\ .
\end{equation}
The RG scaling dimensions $y_{4,2}$ and $y_{4,4}$ of the remaining
perturbations can be associated or to scaling correction exponents,  or to
further crossover exponents, depending whether they correspond to irrelevant
(i.~e. they are negative) or relevant perturbations.

Let us now focus on the case $N_1=2$, $N_2=3$, $N=5$ which is the one relevant for our
study. Results for the three dimensional RG scaling dimensions computed using
different techniques are reported in Tab.~\ref{critexp}, and it is clear that
$y_{4,4}>0$, hence for the symmetry O(2)$\oplus$O(3) to enlarge to O(5) we have
in the general case to tune to zero not only the coefficients of the quadratic
terms $\mathcal{P}_{2,0}$ and $\mathcal{P}_{2,2}$ (as usual at a bicritical
point), but also the coefficient of $\mathcal{P}_{4,4}$. The scaling
of the singular part of the free energy density close to the three dimensional O(5)
multicritical point is thus generically of the form (see, e.~g. Ref.~\cite{PV-02})
\begin{equation}
f_{\mathrm{sing}}=b^{-3}f_{\mathrm{mc}}(b^{y_{2,0}}t, b^{y_{2,2}}g_2, b^{y_{4,4}}g_4)
\end{equation}
where $t$ is the reduced temperature, $g_2$ and $g_4$ are the couplings of the
perturbation $\mathcal{P}_{2,2}$ and $\mathcal{P}_{4,4}$, respectively, and
irrelevant contributions have been neglected for the sake of the simplicity.
By using standard manipulations this scaling form can be written as
\begin{equation}
f_{\mathrm{sing}}=t^{3\nu}f_{\mathrm{mc}}(g_2t^{-\phi_T}, g_4t^{-\phi_Q})\ ,
\end{equation}
where $\nu=1/y_{2,0}$ is the exponent governing the divergence of the critical length in
the three dimensional O(5) model (see Tab.~\ref{crit5}), $\phi_T$ is the
previously introduced crossover exponent, and $\phi_Q=\nu y_{4,4}$ is a further
crossover exponent. 

\begin{table}[t]
\begin{tabular}{lcccc}
\hline\hline
Method            & $y_{2,2}$     & $y_{4,0}$       & $y_{4,2}$       & $y_{4,4}$     \\  \hline
$\varepsilon$ exp\phantom{a}    &  1.832(8)$^a$ &  -0.783(26)$^a$ &  -0.441(13)$^a$ & 0.198(11)$^a$ \\  \hline
$d=3$ exp         &  1.79(5)$^b$\phantom{1}  &  -0.790(15)$^a$ &                 & 0.189(10)$^c$ \\  \hline
MC                &               &  -0.754(7)$^d$  &                 & 0.180(15)$^e$ \\  \hline\hline
\end{tabular}
\caption{Estimates of the three dimensional RG scaling dimensions $y_{2,2}$,
$y_{4,0}$, $y_{4,2}$ and $y_{4,4}$ (for $N_1=2$, $N_2=3$) computed using
different approaches: $\varepsilon$ expansion ($\varepsilon$ exp), fixed dimension
expansion ($d=3$ exp), Monte Carlo (MC). ``a'' stands for Ref.~\cite{CPV-03},
``b'' stands for Ref.~\cite{CPV-02} ($\phi_T=1.40(4)$ is reported in this
reference, we extracted $y_{2,2}$ using $\nu$ for the O(5) model, see
Tab.\ref{crit5}), ``c'' stands for Ref.~\cite{CPV-02-comm}, ``d'' stands for
Ref.~\cite{Hasenbusch-22}, ``e'' stands for Ref.~\cite{HPV-05}.}
\label{critexp}
\end{table}

\begin{table}[t]
\begin{tabular}{ll}
\hline\hline
$\nu$       &  0.7802(6) \\ \hline
$\omega$    &  0.754(7) \\ \hline
$\eta$      &  0.03397(9) \\ \hline
$R_{\xi}^*$ &  0.53691(7) \\ \hline
$U^*$       &  1.069735(25) \\ \hline\hline
\end{tabular}
\caption{Estimates of some critical exponents and universal quantities at the
O(5) critical point, from~\cite{Hasenbusch-22}, where both the ratio
$R_{\xi}=\xi/L$ and the Binder parameter $U$ are computed in the vector
channel, i.~e. starting from the two point function $\langle {\bm s}_{\bm
x}\cdot {\bm s}_{\bm x}\rangle$.}
\label{crit5}
\end{table}

\begin{figure}[b]
\includegraphics[width=0.9\columnwidth, clip]{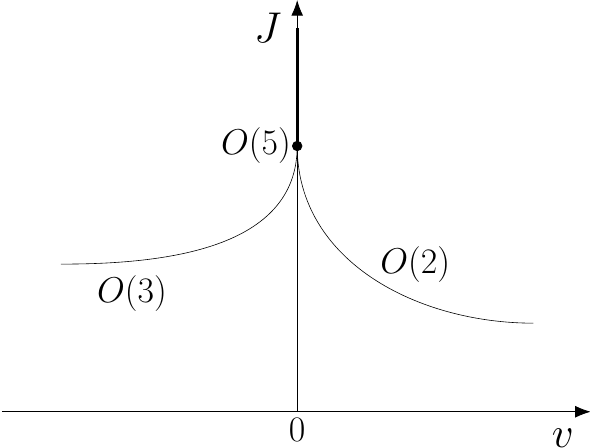}
\caption{Expected phase diagram of the SU(2) gauge model with two flavors of
scalar fields transforming in the fundamental representation, obtained under
the assumption that a standard multicritical O(5) LGW point is present in the
phase diagram (i.~e., that gauge fields are not critical at the transition
and the transition is of LGW type).  Thin lines represent continuous
phase transitions, while the thick line stands for the discontinuous transition
line.}
\label{fig_pd}
\end{figure}

Only if the perturbation $\mathcal{P}_{4,4}$ is not generated by the RG flow,
or if its coefficient is tuned to zero, the standard three-dimensional
bicritical scaling of the free energy density is recovered:
\begin{equation}\label{bicscaling}
f_{\mathrm{sing}}=t^{3\nu}f_{\mathrm{mc}}(g_2t^{-\phi_T})\ .
\end{equation}
In this case a particular surface (identified by the constraint $g_2=0$) is
present in the phase diagram, along which the only relevant RG scaling
dimension is $y_{2,0}=1/\nu\approx 1.282$. Approaching the multicritical point
along any other direction the dominant RG scaling dimension is instead
$y_{2,2}\approx 1.832$. Scaling corrections are governed by the usual exponent
$\omega=-y_{4,0}\approx 0.754$ of the O(5) critical point only if $g_2=0$ and
the perturbation $\mathcal{P}_{4,2}$ is not generated: for $g_2\neq 0$ scaling
corrections with exponent $y_{2,2}-y_{2,0}\approx 0.55$ are
generally expected, which have no counterpart in the critical O(5) model, just
like those associated to the perturbation $\mathcal{P}_{4,2}$.

Let us now come back to the SU(2) gauge model with two flavors of scalar
fields. If we assume that gauge fields are not critical, and that the
critical behavior can be described by a LGW gauge invariant formulation,
then the Hamiltonian in Eq.~\eqref{ham} is equivalent (for what concerns
critical properties) to an O(5) Hamiltonian, corresponding to the $v=0$
Hamiltonian, with a deformation proportional to $v\mathcal{P}_{2,2}$. This can
be easily seen by comparing Eq.~\eqref{trphi2} with the explicit expression of
$\mathcal{P}_{2,2}$ reported, e.~g., in Ref.~\cite{HPV-05}. In the RG flow of
this model the term $\mathcal{P}_{4,4}$, which triggers the instability in the
generic case, is not present, thus we expect for the singular part of the free
energy density a scaling of the form reported in Eq.~\eqref{bicscaling}, and a phase
diagram like the one schematically reported in Fig.~\ref{fig_pd} (it is easily
seen from Eq.~\eqref{bicscaling} that $|J_c(v)-J^*|\sim |v|^{1/\phi_T}$, and
since $\phi_T\approx 1.429>1$ the O(2) and O(3) transition lines approach 
the $v=0$ axis tangentially). Note that, under the above stated assumptions,
the parameter $\gamma$ is expected to be irrelevant. This does not however
exclude the possibility that, for some values of $\gamma$, the transition lines
denoted by O(2) and O(3) in  Fig.~\ref{fig_pd} correspond in fact to
discontinuous transitions.

Some general features of the phase diagram reported in Fig.~\ref{fig_pd} have
been already verified in previous works. In Refs.~\cite{BPV-19a, BPV-19b} it
was shown that the transition of the $v=0$ model is in the O(5) universality
class, indeed the critical exponent governing the divergence of the correlation
length was estimated to be $\nu = 0.775(6)$, cf. Tab.~\ref{crit5}, and also
scaling curves were perfectly consistent with those of the O(5) vector
universality class.  In Ref.~\cite{BFPV-21x} transitions of the O(2) and O(3)
universality classes have been instead identified in numerical simulations for
some $v>0$ and $v<0$ values, respectively. 

In this work we want to test more directly some predictions of
the O(5) multicritical theory. Since the O(5) multicritical point is located at
$v=0$ (where the Hamiltonian display exact O(5) symmetry), and
$\mathcal{P}_{2,2}$ does not mix with O(5) scalars under RG flow, we have
indeed, for small values of $|v|$ and $J\approx J^*$ (where $J^*$ denotes the
value of $J$ at the multicritical point), the scaling 
\begin{equation}\label{bicscaling2}
f_{\mathrm{sing}}(J,v)=b^{-3}
\mathcal{F}_{\mathrm{mc}}(b^{y_{2,0}}(J-J^*), b^{y_{2,2}}v)\ ,
\end{equation}
where corrections to scaling due to irrelevant operators have been neglected
for the sake of the simplicity.  In particular, by fixing $J=J^*$ and varying
the parameter $v$, it is possible to extract the critical exponent
$\nu'=1/y_{2,2}$, which to the best of our knowledge has never been estimated so
far by Monte Carlo methods, but only in $\varepsilon$ and fixed dimension
expansions, see Tab.~\ref{critexp}. 
For the analyses of the following section
it will be useful to note that, due to the exact O(5) symmetry of the model when
$v=0$, the scaling function $\mathcal{F}_{\mathrm{mc}}(x,v=0)$ is the same
scaling function of the O(5) model (obviously up to non-universal rescalings).

\section{Numerical results}
\label{results}

\subsection{Simulation algorithm}

To sample the distribution $e^{-H}/Z$, with $H$ defined in Eq.~\eqref{ham}, it
is convenient to use a combination of heat-bath, microcanonical and Metropolis
update schemes.  Link SU(2) variables are updated using standard heat-bath
\cite{KP-85} and microcanonical \cite{Creutz-87} algorithms.  A similar
strategy has been adopted for the complex scalar field $\Phi^{af}$: to
ergodically sample these variables we perform a Metropolis update
\cite{MRRTT-53}, which rotates two randomly chosen elements of $\Phi_{\bm
x}^{af}$. Denoting these elements by $\phi_1$ and $\phi_2$, the proposed trial
state is  
\begin{equation}
\begin{aligned}
\phi'_1 &= \cos\theta_1 e^{i\theta_2} \phi_1 + \sin\theta_1 e^{i\theta_3}\phi_2\\
\phi'_2 &= 
   -\sin\theta_1 e^{i\theta_2}\phi_1 + \cos\theta_1 e^{i\theta_3}\phi_2 \,,
\end{aligned}
\label{metro_rotation}
\end{equation}
where the angles $\theta_i$ are sampled from the uniform distribution in
$[-\alpha,\alpha]$ (to ensure detailed balance), and the value of $\alpha$ has
been fixed in such a way that the acceptance probability of the Metropolis
filter is not smaller than 30\%. To reduce autocorrelations we also use a different
Metropolis update: let us define $\Phi'_{\bm x}$ by 
\begin{equation}
  \Phi'_{\bm x} = \frac{2 \mathrm{Re}\Tr(\Phi^\dagger_{\bm x} S_{\bm x})}
       {\Tr( S^\dagger_{\bm x} S_{\bm x})}S_{\bm x} 
- \Phi_{\bm x}\,,
\label{generalized-overrelaxed}
\end{equation}
where $S_{\bm x}$ is the matrix
\begin{equation}
    S_{\bm x} = \sum_\mu 
    (U_{{\bm x}, \mu} \Phi_{{\bm x}+\hat{\mu}} +
    U^\dagger_{{\bm x}-\hat{\mu}, \mu} \Phi_{{\bm x}-\hat{\mu}}) \,.
\end{equation}
Since the transformation $\Phi_{\bm x}\to \Phi'_{\bm x}$ is involutive,
$\Phi'_{\bm x}$ is a legitimate trial state for a Metropolis update, moreover
for $v=0$ it is easily seen that the transformation $\Phi_{\bm x}\to \Phi'_{\bm
x}$ constitutes a microcanonical update of the scalar fields, hence it is
always accepted.  For $v\neq 0$ the trial state $\Phi'_{\bm x}$ is accepted or
rejected using a Metropolis test, which turns out to have a large acceptance
rate for all the cases studied in this paper. For this reason we call this
update step pseudo-microcanonical update.

We call lattice iteration a sequence of 11 update sweeps on all the lattice
variables: for link variables an heat-bath update is followed by ten
microcanonical updates, while for scalars a Metropolis update with trial state
Eq.~\eqref{metro_rotation} is followed by ten pseudo-microcanonical updates.
Measures were performed after every lattice iteration, and the gathered
statistic has been of the order of $10^6$ iterations in
all the cases.  Data have been analyzed by using standard jackknife and
blocking techniques.

\subsection{Observable and Finite Size Scaling}

The observables that we use to study critical correlations of $Q_{\bm x}$ (see
Eq.~\eqref{Q}) and $Y_{\bm x}$ (see Eq.~\eqref{Y}) can be built starting from
their respective two point functions:
\begin{equation}
\begin{aligned}
G_Q({\bm x}-{\bm y})&=\langle\mathrm{Tr}Q_{\bm x}Q_{\bm y}\rangle\ ,\\
G_Y({\bm x}-{\bm y})&=\langle\mathrm{Tr}Y_{\bm x}^{\dag}Y_{\bm y}\rangle\ .
\end{aligned}
\end{equation}
The second moment correlation lengths $\xi_Q$ and $\xi_Y$ are defined by 
\begin{equation}
\xi_{\#}^2 = \frac{1}{4 \sin^2 (\pi/L)} \frac{\widetilde{G}_{\#}({\bm 0}) -
  \widetilde{G}_{\#}({\bm p}_m)}{\widetilde{G}_{\#}({\bm p}_m)}\,,
\end{equation}
where $\#$ stands for $Q$ or $Y$, ${\bm p}_m = (2\pi/L,0,0)$, and
$\widetilde{G}_{\#}({\bm p})=\sum_{{\bm x}} e^{i{\bm p}\cdot {\bm x}}
G_{\#}({\bm x})$ is the Fourier transform of the two point function $G_{\#}(\bm
x)$.  Since the finite size scaling (FSS) of RG invariant quantities is
particularly simple, we define the ratios 
\begin{equation}
R_{\#}=\xi_{\#}/L\ ,
\end{equation}
the Binder parameter $U_Q$
\begin{equation}
U_{Q} = \frac{\langle \mu_2^2\rangle}{\langle \mu_2 \rangle^2} \,, \qquad
\mu_2 = \frac{1}{L^6}  
\sum_{{\bm x},{\bm y}} {\rm Tr}\,Q_{\bm x} Q_{\bm y}\,,
\end{equation}
and the analogously defined Binder parameter $U_Y$, all of which are invariant
RG functions. 

Using Eq.~\eqref{bicscaling2} it is simple to verify (see, e.~g. \cite{PV-02})
that the FSS of any RG invariant quantity $M$ close to the O(5) multicritical
point is of the form\footnote{Up to scaling corrections which vanish for large
values of $L$ with a different exponent if $v=0$ or $v\neq 0$.}
\begin{equation}\label{scaling}
M(J,v)=\mathcal{M}(X)\ ,
\end{equation}
where
\begin{equation}
X=\left\{\begin{array}{ll} (J-J^*)L^{y_{2,0}} & v=0 \\
vL^{y_{2,2}} & v\neq 0 \end{array}\right.\ ,
\end{equation}
and we used the fact that
\begin{equation}
y_{2,2}\approx 1.832 > 1.282 \approx y_{2,0}\ .
\end{equation}

\begin{figure}[t]
\includegraphics[width=0.9\columnwidth, clip]{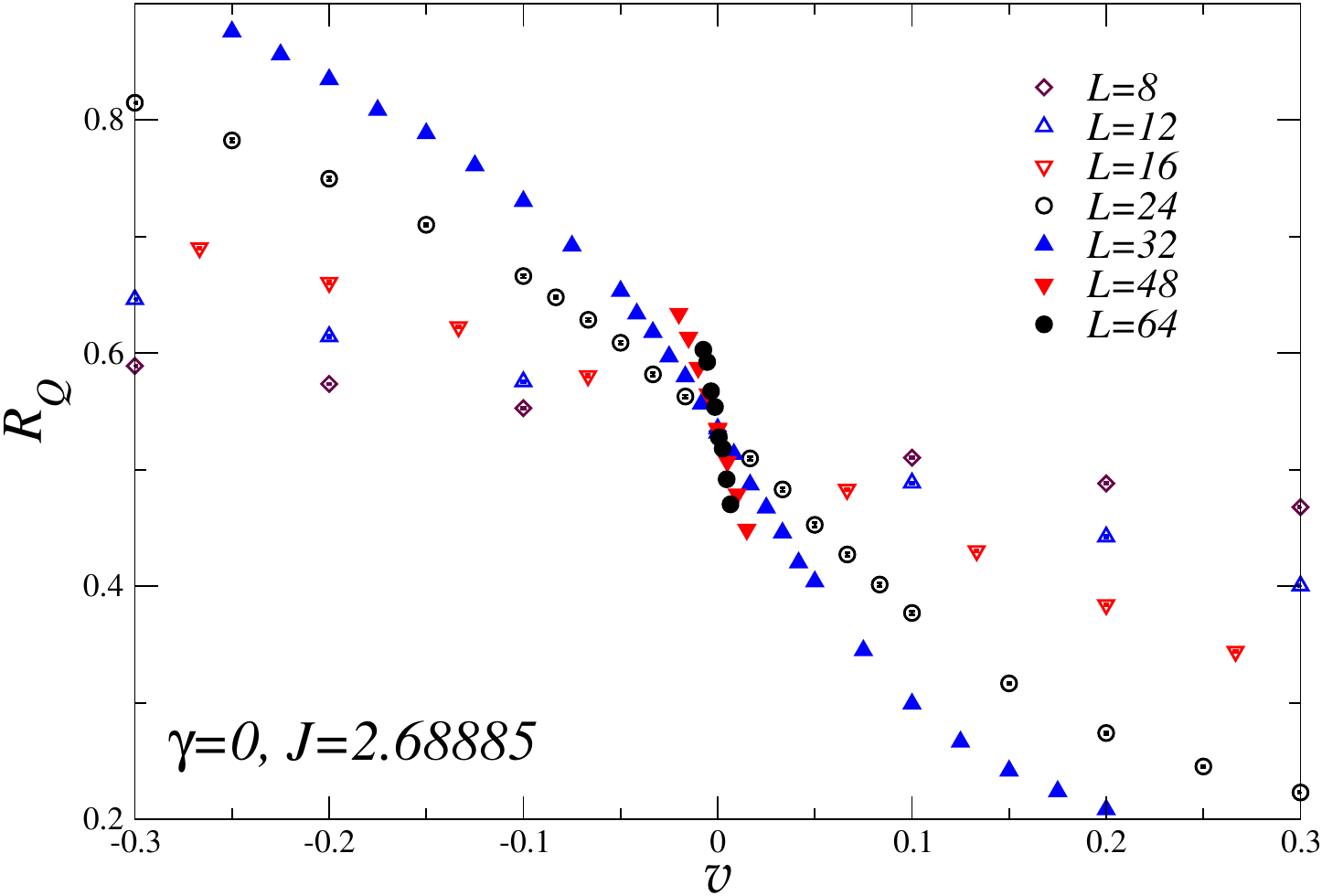}\\
\includegraphics[width=0.9\columnwidth, clip]{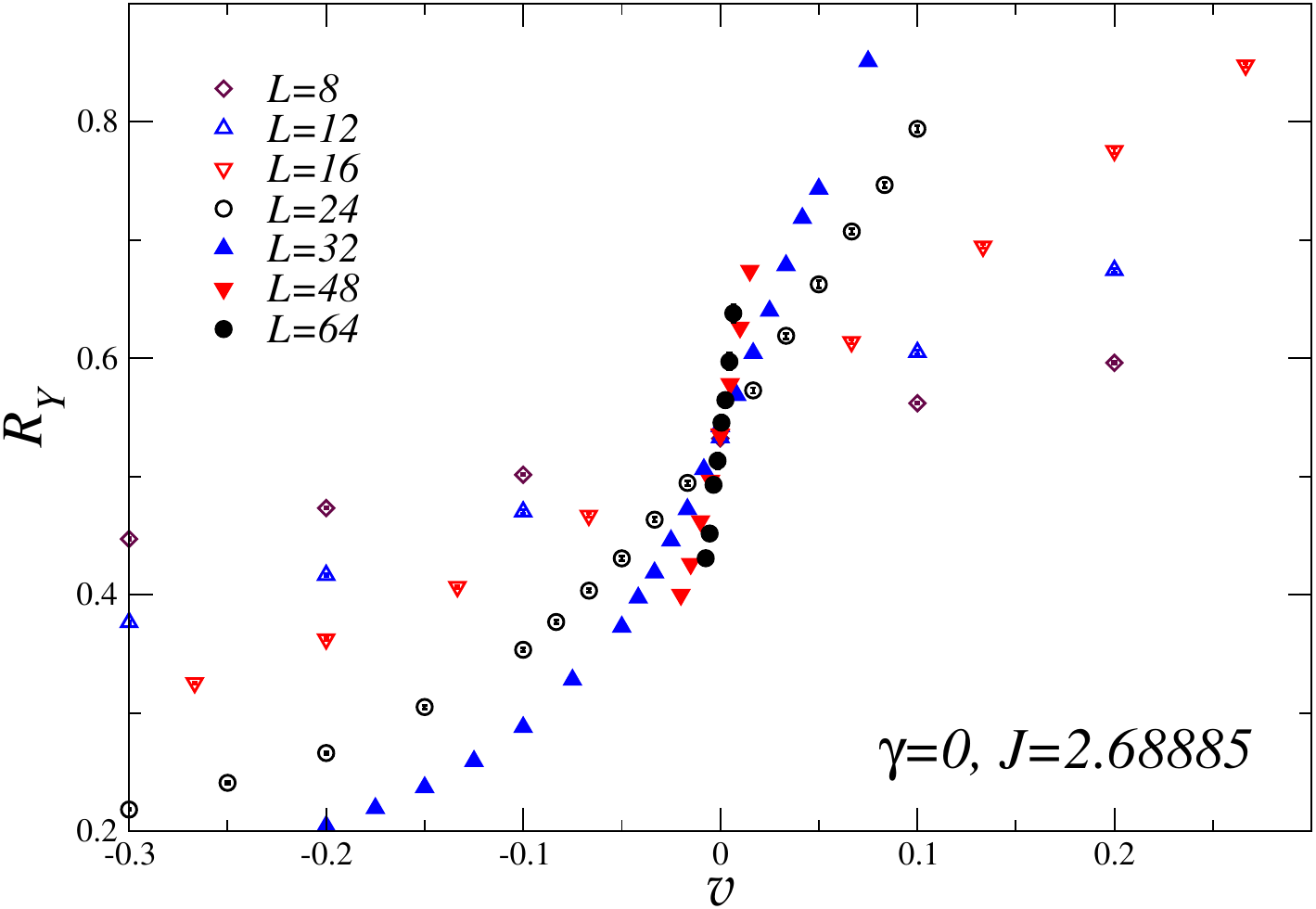}
\caption{Behavior of $R_Q$ (upper panel) and $R_Y$ (lower panel) as a function
of $v$ for $\gamma=0$ and $J=2.68885$.}
\label{fig_R}
\end{figure}

\subsection{Results}
 
\begin{figure}[b]
\includegraphics[width=0.9\columnwidth, clip]{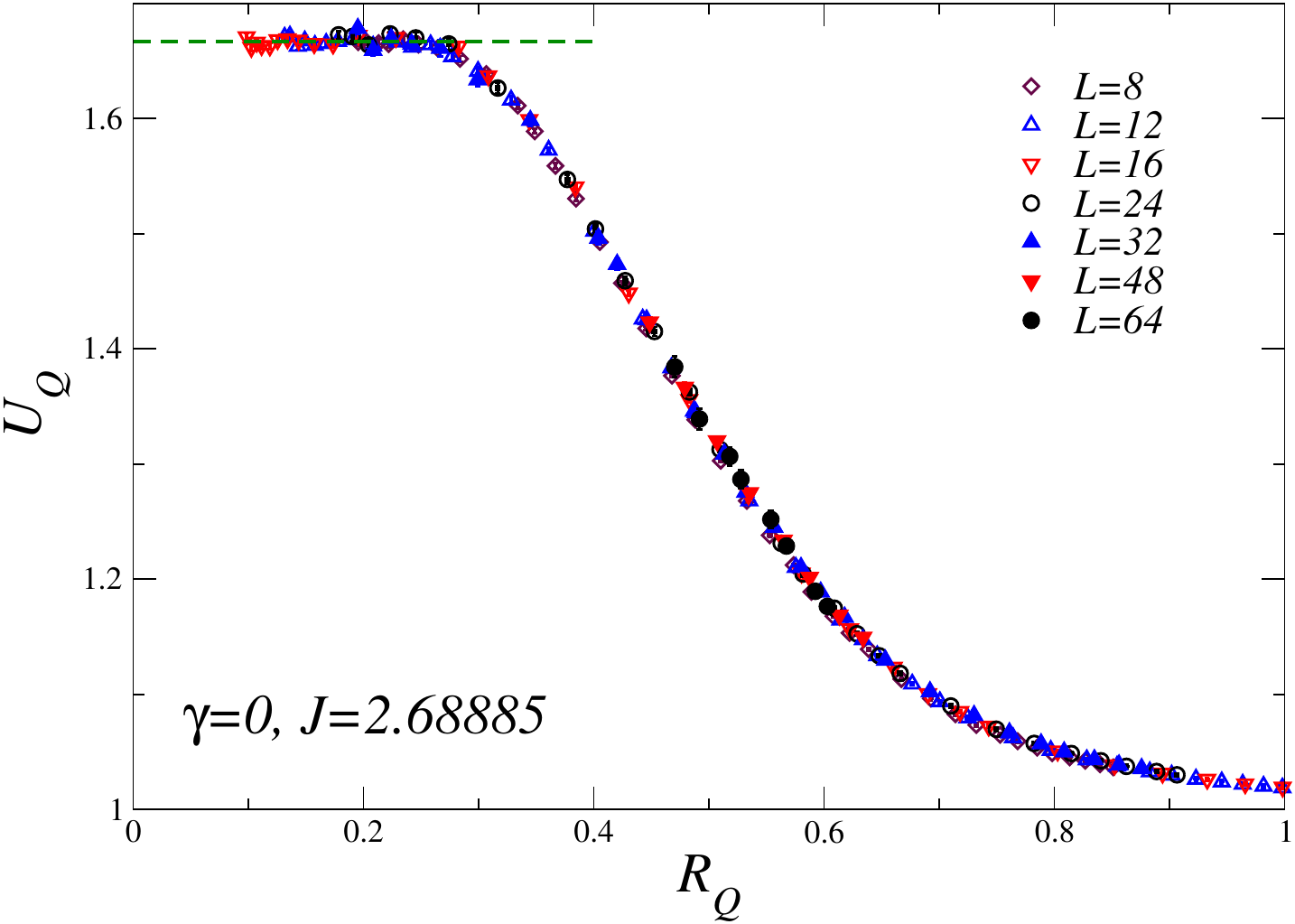}\\
\includegraphics[width=0.9\columnwidth, clip]{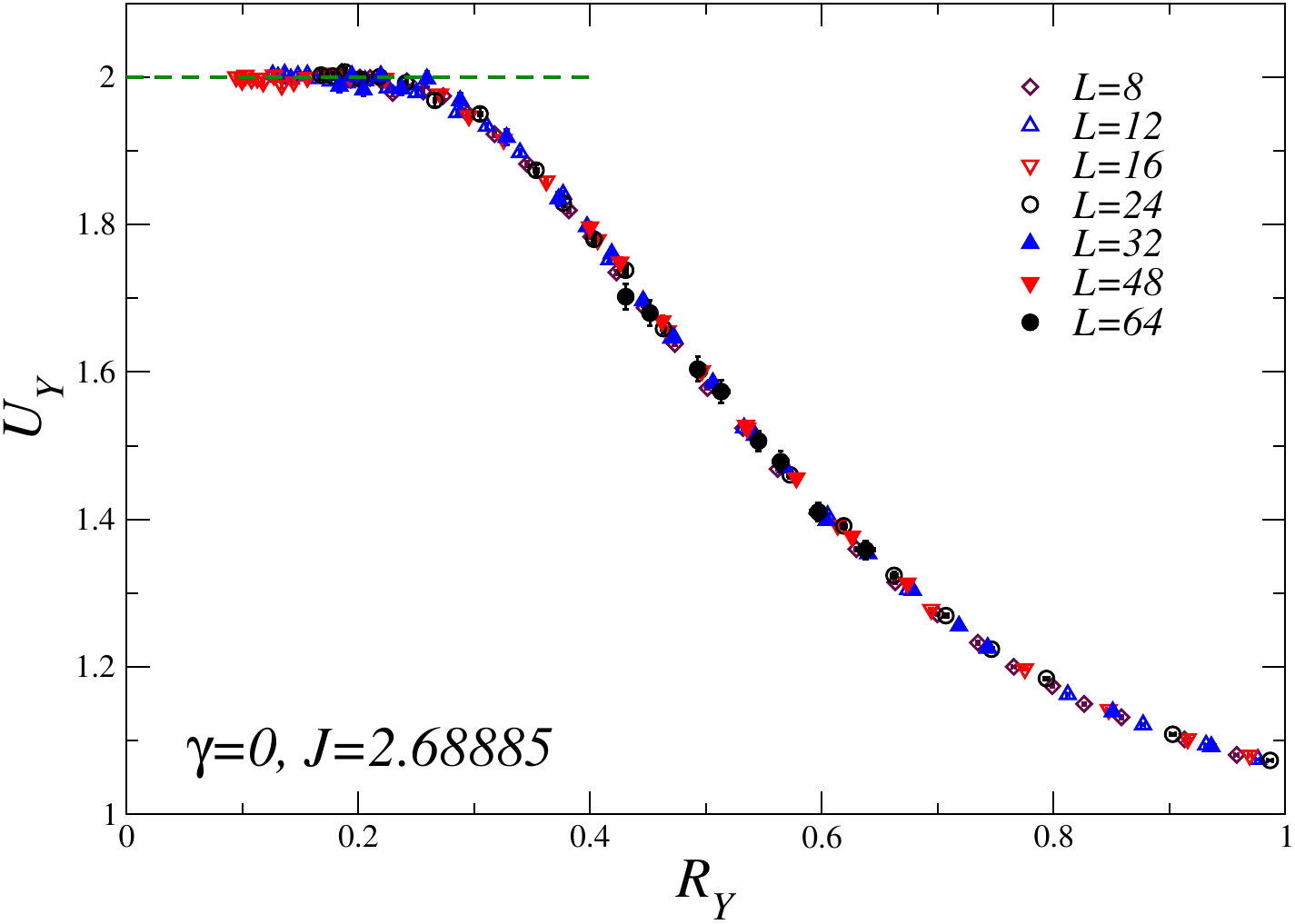}
\caption{Behavior of $U_Q$ as a function of $R_Q$ (upper panel) and of $U_Y$ as a
function of $R_Y$ (lower panel) for $\gamma=0$ and $J=2.68885$. The horizontal
dashed line denote the analytically known low temperature values of $U_Q$ and
$U_Y$ expected from O(3) and O(2) symmetry, respectively, 
which are $(N+2)/N$ for $N=3$ and $N=2$, respectively.}
\label{fig_U_R}
\end{figure}

To verify the multicritical O(5) scaling we use mainly simulations at
$\gamma=0$, fixing $J=J^*(\gamma=0)$ and varying the value of the parameter
$v$, for lattice sizes going from $L=8$ to $L=64$. The multicritical coupling
$J^*(\gamma)$ has been determined in Refs.~\cite{BPV-19a, BPV-19b} for some
values of the gauge coupling $\gamma$, obtaining in particular 
\begin{equation}
\begin{aligned}
J^*(0)&=2.68885(5)\ ,\\
J^*(2)&=1.767(1)\ ,\\
J^*(-2)&=3.794(2)\ .
\end{aligned}
\end{equation}
To check that the $\gamma$
parameter is irrelevant we also performed some simulations at $\gamma=2$, as
will be discussed in the following. 

The measured values of $R_Q$ and $R_Y$, as a function of $v$, are reported in
Fig.~\ref{fig_R} for simulations performed at $\gamma=0$ and $J=J^*(0)$. In
both the cases a crossing point at $v=0$ can be clearly seen, as expected at a
continuous phase transition. The different behavior of the $R_Q$ and $R_Y$ data
is moreover consistent with the phase diagram in Fig.~\ref{fig_pd}: the
variables $Q_{\bm x}$ displays long range order for $v<0$, while the variables
$Y_{\bm x}$ displays long range order for $v>0$. The transition at $v=0$,
$J=J^{*}$ is thus a continuous transition between two different ordered phases.
The behaviors or $U_Q$ and $U_Y$ as function of $v$ (not shown), confirm this picture. 

Before discussing the determination of $y_{2,2}$, let us note that from
Eq.~\eqref{scaling} it immediately follows that, at a continuous phase
transition, data for $U_Q$ computed on different lattice sizes should
asymptotically collapse on a single curve when plotted as a function of $R_Q$
(and analogously for $U_Y$ and $R_Y$).  These collapse plots are reported in
Fig.~\ref{fig_U_R}, and they show that practically no scaling corrections are
present. In the same plots we also report (as horizontal dashed lines) the
FSS low temperature values of $U_Q$ and $U_Y$ expected from O(3) and O(2)
symmetry. At a LGW transition the gauge invariant bilinear operators $Q_{\bm
x}$ and $Y_{\bm x}$ correspond to the fundamental fields to be used in the
effective Lagrangian (see, e.~g., \cite{BPV-24-rev}), hence the asymptotic
values of their Binder parameters for $R_Q,R_Y\to 0$ coincide with those of the
Binder parameter of the vector correlation functions in O($N$) models; their
values are thus expected to be $(N+2)/N$ for $N=3$ and $N=2$, for
$Q_{\bm x}$ and $Y_{\bm x}$, respectively.

\begin{table}[t]
\begin{tabular}{lccc}
\hline\hline
Obs.  & $v_c$     & $\nu'$    & Obs$^*$   \\  \hline
$U_Q$       & 0.0005(5) & 0.543(7)\phantom{0} & 1.28(1)\phantom{00}    \\  \hline
$U_Y$       & 0.0000(5) & 0.547(6)\phantom{0} & 1.52(1)\phantom{00}   \\  \hline
$R_Q$       & 0.0000(5) & 0.545(10)           & 0.538(5)\phantom{0}  \\  \hline
$R_Y$       & 0.0005(5) & 0.545(4)\phantom{0} & 0.545(10) \\ \hline\hline
\end{tabular}
\caption{Numerical results of unbiased fits to data of the observable Obs.,
performed by fitting both the critical coupling $v_c$ (i.~e. without fixing
$v_c=0$) and the values of the observable at the critical point, denoted by
Obs$^*$.}
\label{fitunbias}
\end{table}

\begin{table}[t]
\begin{tabular}{lc}
\hline\hline
Observ.  & $\nu'$     \\  \hline
$U_Q$    & 0.542(4)\phantom{0}  \\  \hline
$U_Y$    & 0.543(4)\phantom{0}  \\  \hline
$R_Q$    & 0.550(10)   \\  \hline
$R_Y$    & 0.544(3)\phantom{0}  \\ \hline\hline
\end{tabular}
\caption{Numerical results of biased fits, performed by fixing $v_c=0$, 
$U_Q^*=\frac{25}{21}U^*$, $U_Y^*=\frac{10}{7}U^*$, $R_Q^*=R_Y^*=R_{\xi}^*$, 
where $U^*$ and $R_{\xi}^*$ denote the critical values obtained from vector correlations
at the standard O(5) critical point (see Tab.\ref{crit5}).}
\label{fitbias}
\end{table}

\begin{figure}[t]
\includegraphics[width=0.9\columnwidth, clip]{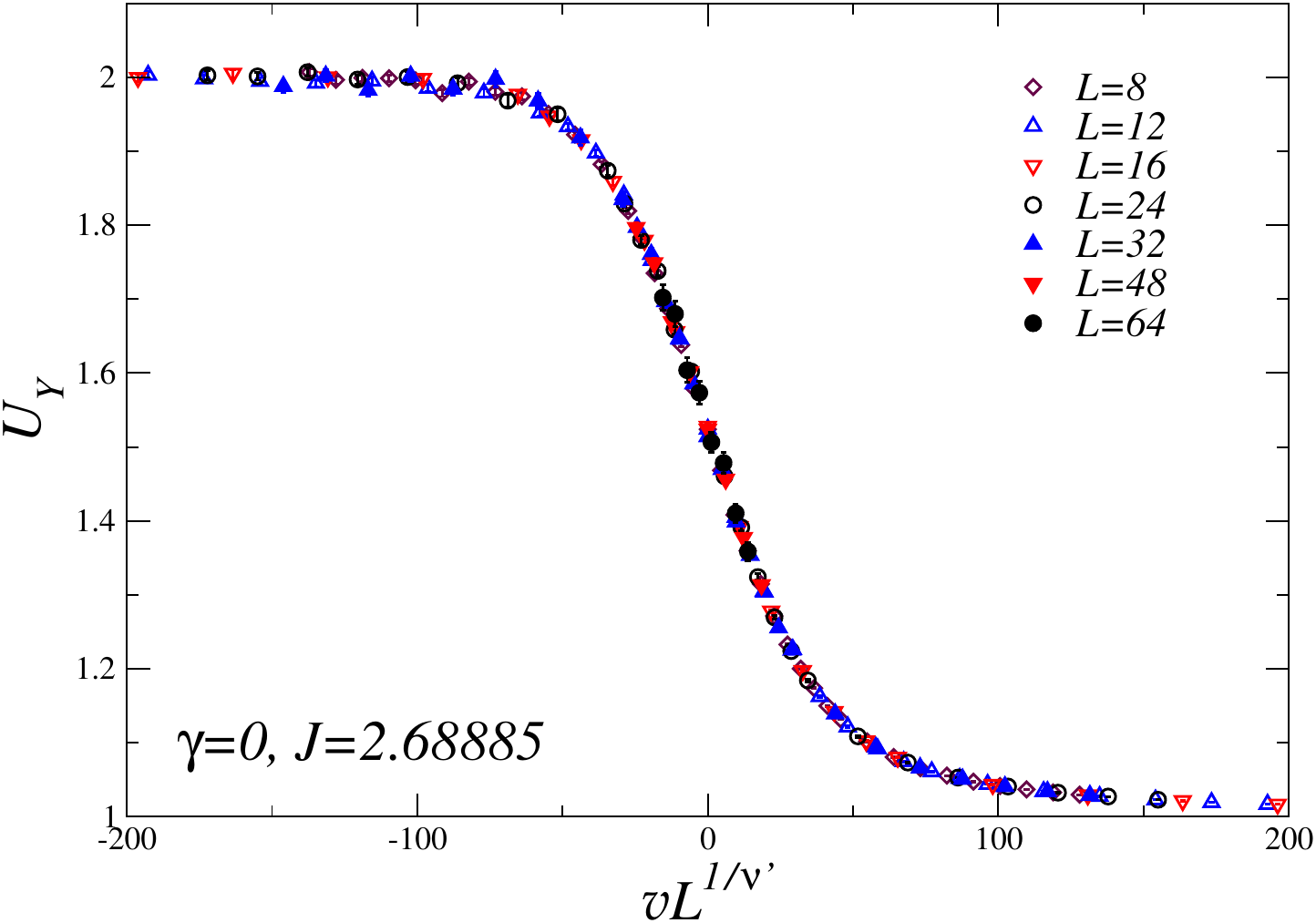}\\
\includegraphics[width=0.9\columnwidth, clip]{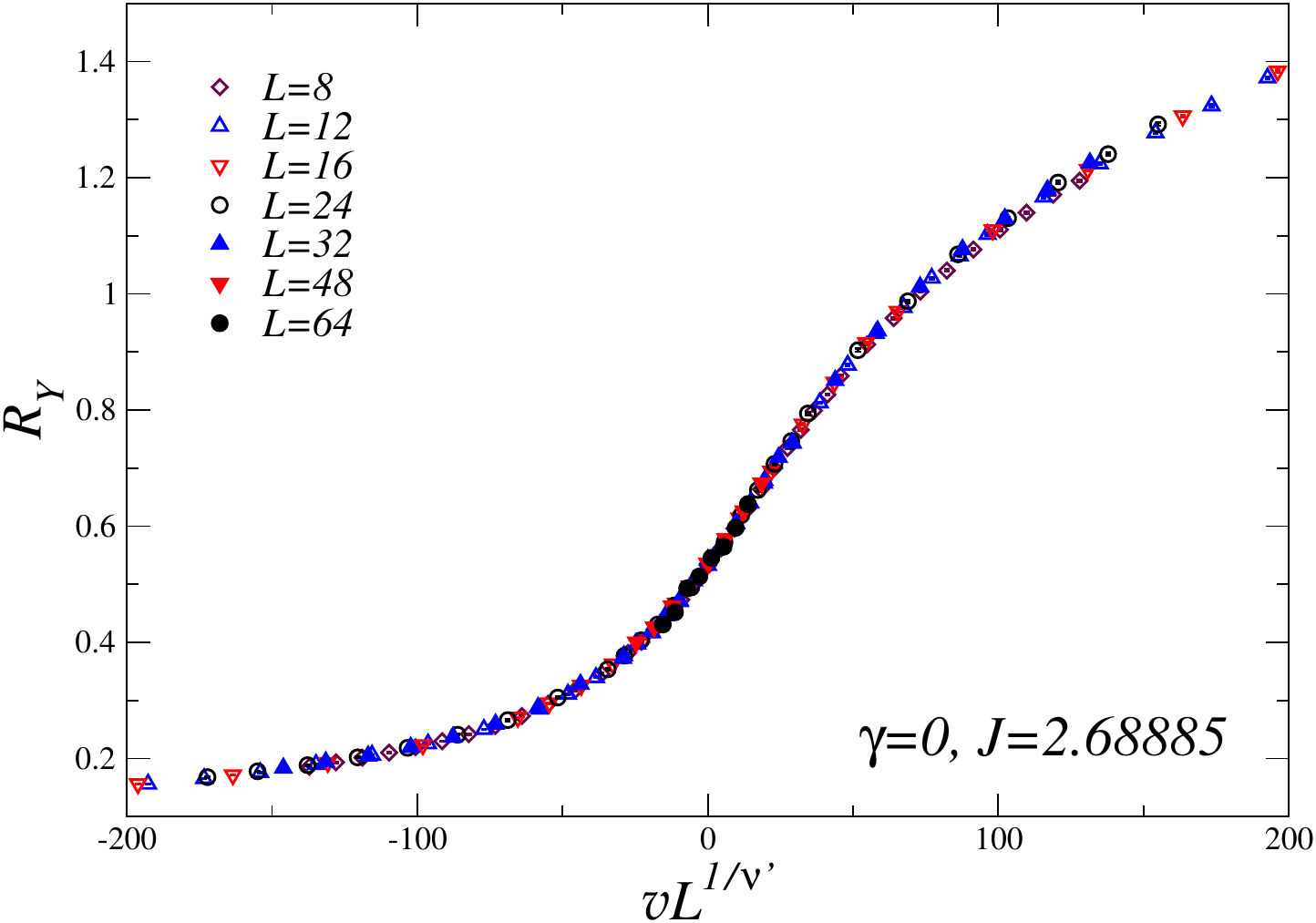}
\caption{Finite size scaling curves of $U_Y$ (upper panel) and $R_Y$ (lower panel),
computed using $\nu'=0.544$.}\label{O5scal}
\end{figure}

To extract the value of $y_{2,2}$ we use the FSS relation in
Eq.~\eqref{scaling}, considering a Taylor series expansion of the scaling
function $\mathcal{M}$. Since our data are not precise enough to clearly
identify scaling corrections we tested several values of the exponent governing
scaling corrections, in the range $[0,1]$. Differences between the results of
different fits (e.~g., different polynomial approximations, different fit ranges,
different values of the scaling correction exponent) have been used to estimate
the systematic uncertainties of the adopted procedure.

We first tried completely unbiased fits, which do not even assume the
transition to happen at $v=0$, whose results are reported in
Tab.~\ref{fitunbias}. The fact that the critical values of $v$ are compatible
with zero in all the cases is a consistency check for the value of $J^*(0)$ computed in
Refs.~\cite{BPV-19a, BPV-19b}, moreover the values at the critical point of the
observables are consistent with expectations based on O(5) symmetry. It has indeed
been shown in \cite{BPV-19a, BPV-19b} that the following relations hold at
criticality in the presence of O(5) symmetry:
\begin{equation}
U_Q^*=\frac{25}{21}U^*\ ,\quad
U_Y^*=\frac{10}{7}U^*\ ,\quad
R_Q=R_Y=R_{\xi}\ ,
\end{equation}
where $U^*$ and $R_{\xi}^*$ are the critical values computed in the O(5) model,
reported in Tab.~\ref{crit5}. We thus expect the values
\begin{equation}
\begin{aligned}\label{eqfixbias}
&U_Q^*\approx 1.27349\ ,\quad \\
&U_Y^*\approx 1.52819\ ,\quad \\
&R_Q^*=R_Y^*\approx 0.53691\ , \\
\end{aligned}
\end{equation}
which are indeed consistent with those reported in Tab.~\ref{fitunbias}.  

Having checked that we have a good control on the systematical errors of the
fit procedure, we then biased the fit by fixing $v_c=0$ and the critical values
in Eq.~\eqref{eqfixbias}. The results obtained for $\nu'=1/y_{2,2}$ are
reported in Tab.~\ref{fitbias}, which are consistent with the values obtained
from the unbiased fits (compare with Tab.~\ref{fitunbias}) but with smaller
errors.  Given the nice agreement within the different estimates, we report as
our final estimate the value
\begin{equation}
\nu' = 0.544(3)\ ,
\end{equation}
corresponding to 
\begin{equation}
y_{2,2} = 1.838(10)\ .
\end{equation}
This value is perfectly consistent with the value 1.832(8) estimated by means
of the $\varepsilon$-expansion~\cite{CPV-03}, and it appears to be the first Monte
Carlo determination of this critical exponent. In Fig.~\ref{O5scal} we report
some collapse plots to show the quality of the scaling obtained
by using $\nu' = 0.544$.

\begin{figure}[t]
\includegraphics[width=0.9\columnwidth, clip]{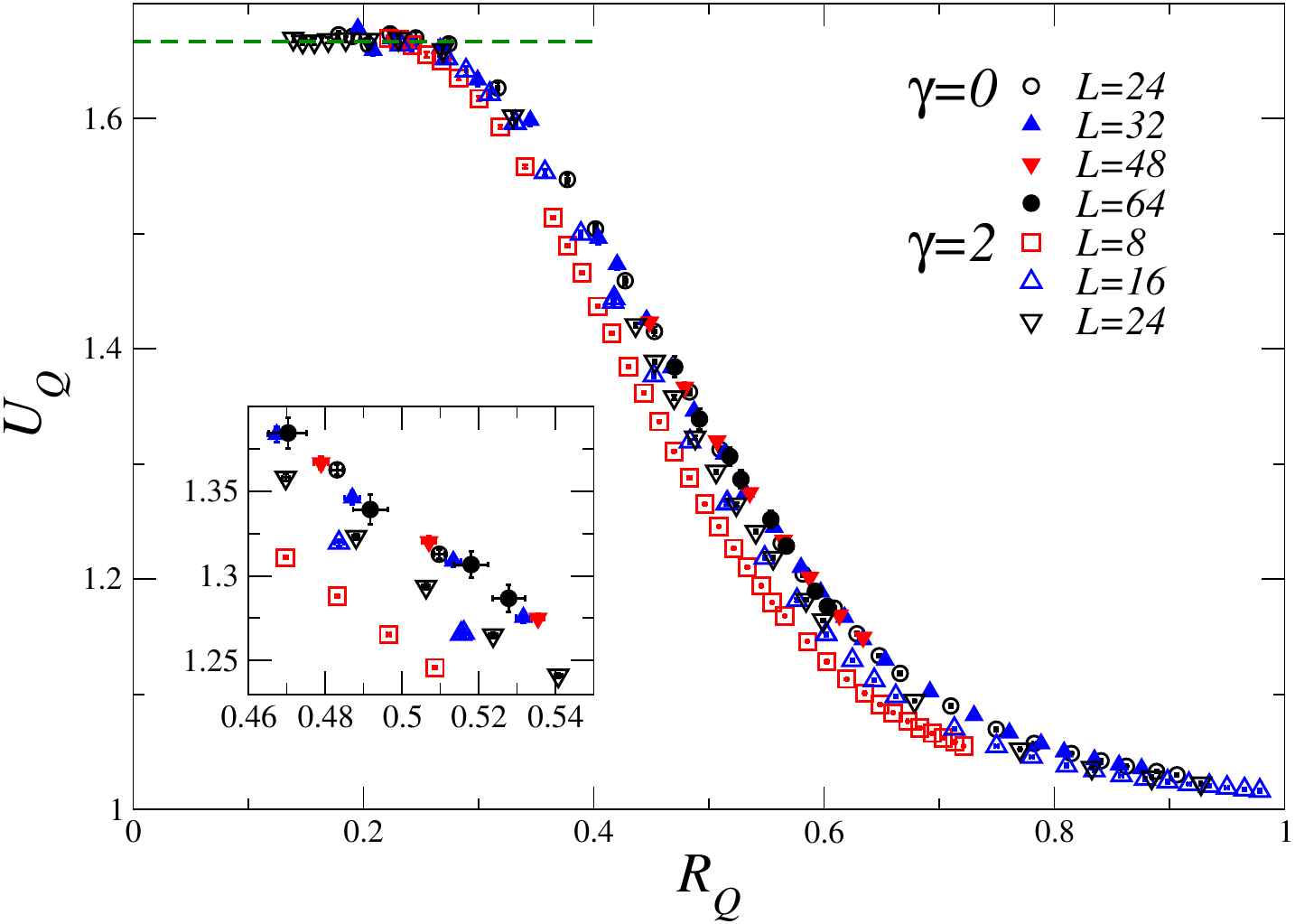}
\caption{Universal scaling curve obtained by plotting $U_Q$ as a function of $R_Q$. Data
obtained from simulations performed using $\gamma=2$ and $J=J^*(2)$ are
compared with data previously shown in Fig.~\ref{fig_U_R}, computed using
$\gamma=0$ and $J=J^*(0)$. Larger scaling corrections are present in the
$\gamma=2$ case, but the approach to the same scaling curve is clear. 
}\label{O5gamma2}
\end{figure}

In the previous discussion we have always assumed 
the critical behavior to be encoded by an effective Hamiltonian written in
terms of local and gauge invariant composite operators, in particular
independent of the gauge fields.  Since $\gamma$ parametrizes the
self-interaction of the gauge fields, if the previous assumption is correct
$\gamma$ has to be an irrelevant coupling. To verify that this is indeed the case
we have also performed simulations for a value of $\gamma$ different from zero,
and in particular for $\gamma=2$. Results obtained for $U_Q$ as a function of
$R_Q$ when using $\gamma=2$ and $J=J^*(2)$ are shown in Fig.~\ref{O5gamma2},
together with the results obtained for $\gamma=0$. Since both $R_Q$ and $U_Q$
are independent of nonuniversal normalizations, the asymptotic scaling of the
two curves should be the same if $\gamma$ is an irrelevant parameter. Data
obtained from simulations at $\gamma=2$ have larger scaling corrections than at
$\gamma=0$, but it is quite evident that both datasets are converging to the
same scaling curve.

\section{Conclusions}
\label{concl}

In this work we numerically investigated the multicritical behavior of the
three dimensional system in which a SU(2) gauge field interacts with two
degenerate flavors of scalar fields. This system is symmetric under
O(2)$\oplus$O(3) global transformations, and its phase diagram displays two
lines of continuous transitions of the O(2) and O(3) universality classes,
merging at  a bicritical point where the global symmetry enlarges to O(5). 
This symmetry enlargement does not happen in generic systems
since it requires the tuning of a further parameter, due to the presence of a
further relevant operator. The existence of this relevant operator is however
prevented in this case by the gauge symmetry. 

All predictions of the multicritical $\phi^4$ theory with 
O(2)$\oplus$O(3) symmetry have been nicely verified in the model considered, 
and in particular we obtained the first Monte Carlo estimate of the critical 
exponent associated to the spin 2 quadratic perturbation $\mathcal{P}_{2,2}$:
\begin{equation}
y_{2,2} = 1.838(10)\ ,
\end{equation}
which is in excellent agreement with the estimate 1.832(8) coming from
$\varepsilon$-expansion~\cite{CPV-03}. These results provide the first evidence
of the fact that, in gauge theories, multicritical phenomena emerging from the
crossing of independent LGW transition lines can
be described by standard LGW $\phi^4$ multicritical theory.

We finally note a possible interesting connection of the results obtained in
this paper with some numerical results obtained in models that are very
different from that considered here. We have explicitly seen that the SU(2)
gauge symmetry plays a fundamental role to protect the symmetry enlargement
O(2)$\oplus$O(3)$\to$O(5), preventing the appearance of the relevant
perturbation $\mathcal{P}_{4,4}$.  Since several studies of models related to
high-$T$ superconductivity \cite{Zhang-97, AH-99, ZHAHA-99, AH-00, DHZ-04} and
to deconfined criticality \cite{TH-05, SF-06, NCSOS-15, NSCOS-15-bis, TS-20,
ZHZH-24, DLGL-24, DES-24} have found signals of emerging O(5) symmetry, it
seems natural to guess that some of these results may be interpreted as due to
the emergence of a SU(2) gauge symmetry coupled to bosonic effective degrees of
freedom.  Obviously it would be unrealistic to expect this to happen in all the
cases, and different explanations are possible (see, e.~g.,
\cite{Senthil:2023vqd} for a recent review of some theoretical models of deconfined
criticality), however this is a possible further mechanism that can be used to interpret 
numerical results.
Also the presence of unusual scaling corrections
\cite{JNCW-08, Sandvik-10, BDA-10, Kaul-11, ZHDKPS-13, HSOMLWTK-13} could be
easily interpreted by the peculiarities of multicritical scaling, in which the
values of both the dominant relevant scaling dimension and of the leading
scaling correction depend on the direction along which the multicritical point
is approached. Note however that most of these models are quantum models,
displaying zero temperature quantum phase transitions, while the model studied
in the present work is a classical model, with a finite temperature phase
transition. To understand whether something different could happen in a quantum
model (e.~g. due to the anisotropy of the quantum to classical mapping) surely
requires further investigation.

\acknowledgments

It is a pleasure to thank Ettore Vicari for useful discussions.
The authors acknowledge support from project PRIN 2022 ``Emerging
gauge theories: critical properties and quantum dynamics''
(20227JZKWP). Numerical simulations have been performed on the CSN4
cluster of the Scientific Computing Center at INFN-PISA.

\end{document}